\documentclass[onecolumn,amsmath,amssymb,amsfonts,a4paper,aps,prd,10pt]{revtex4}
\usepackage{graphicx}
\usepackage{epstopdf}
\newcommand{\nc}{\newcommand}
\usepackage{nicefrac}
\usepackage{amssymb}
\usepackage{amsmath}
\usepackage{mathrsfs}
\usepackage{graphics}
\usepackage{epsfig}
\usepackage{graphicx}
\usepackage{subfigure}
\usepackage{bm}
\usepackage{nicefrac}

\def\be{\begin{equation}}
\def\ee{\end{equation}}
\def\bea{\begin{eqnarray}}
\def\eea{\end{eqnarray}}
\def\ba{\begin{eqnarray}}
\def\ea{\end{eqnarray}}

\def\tr{{\rm tr}}
\def\mK{{\mathcal K}}

\nc{\D}{\overline{\mbox{D3}}}

\nc{\ga}{\gamma} \nc{\tnu}{\tilde{\nu}} \nc{\tmu}{\tilde{\mu}}

\newcommand{\f}[2]{\frac{#1}{#2}}

\input{epsf.sty}
\begin{document}

\title{

Stability of the Self-accelerating Universe in Massive Gravity}

\author{Nima Khosravi$^{(1)}$, Gustavo Niz$^{(2,3)}$, Kazuya Koyama$^{(2)}$, and Gianmassimo Tasinato$^{(2)}$}
\affiliation{%
~ \\$^{(1)}$ Cosmology Group, African Institute for Mathematical
Sciences, Muizenberg, 7945, South Africa,\vspace{.2cm}\\South
African Astronomical Observatory, Observatory Road, Observatory,
Cape Town, 7935, South Africa,\vspace{.2cm}\\Department of
Mathematics and Applied Mathematics, University of Cape Town,
Rondebosch, Cape Town, 7700, South Africa,
 \vspace{.5cm} \\
 $^{(2)}$ Institute of Cosmology $\&$ Gravitation, University of Portsmouth,
\\\hskip0.2cm 
 Portsmouth, PO1 3FX, United Kingdom, \vspace{.5cm}\\
$^{(3)}$ Departamento de F\'{\i}sica, Universidad de Guanajuato,\\
 DCI, Campus Le\' on, C.P. 37150, Le\' on, Guanajuato, M\' exico.
}%

\begin{abstract}
We study linear perturbations around time dependent spherically
symmetric solutions in the $\Lambda_3$ massive gravity theory, which
self-accelerate in the vacuum. We find that the dynamics of the
scalar perturbations depend on the choice of the fiducial metric for
the background solutions. For particular choice of fiducial metric
there is a symmetry enhancement, leaving no propagating scalar
degrees of freedom at  linear order in perturbations. In contrast,
any other choice propagates a single scalar mode. We find that the
Hamiltonian of this scalar mode is unbounded from below for all
self-accelerating solutions, signalling an instability.
\end{abstract}

 \maketitle


\section{Introduction and Summary}

It is a standing question whether the $\Lambda$CDM model is the
correct description of the recent cosmic acceleration. Modified
gravity models, such as massive gravity, may provide an alternative
description to the cosmological constant scenario, where the
background solution mimics precisely an isotropic and homogeneous
background driven by a cosmological constant. Therefore, in order to
discriminate between General Relativity (GR) and modified gravity,
it is important to understand the evolution of perturbations on
these backgrounds.

 Fierz and Pauli (FP), back in 1939, started
 the theoretical
 study
 of massive
 gravity from a field theory perspective \cite{Fierz:1939ix}.
  They considered a mass term for linear gravitational perturbations, which is uniquely determined by requiring the absence of ghost degrees of freedom. The mass term breaks the gauge (diffeomorphism) invariance of GR, leading to a classical graviton with five degrees of freedom, instead of the two found in GR. There have been intensive studies into what happens beyond the linearized theory of FP. In 1972, Boulware and Deser (BD) found a scalar ghost mode at the nonlinear level, the so called sixth degree of freedom in the FP theory \cite{Boulware:1973my}.  This issue has been re-examined using an effective field theory approach,  where gauge invariance is restored by introducing St\"uckelberg fields \cite{ArkaniHamed:2002sp}. In this language, the St\"uckelberg fields
physically play the role of the additional scalar and vector
graviton polarizations. They acquire nonlinear interactions which
contain more than two time derivatives, signaling the existence of a
ghost \cite{ArkaniHamed:2002sp}. In order to construct a consistent
theory, nonlinear terms should be added to the FP model, which are
tuned to remove the ghost order by order in perturbation theory.
Interestingly, this approach sheds light on another famous problem
with FP massive gravity; due to contributions of the scalar degree
of freedom, solutions in the FP model do not continuously connect to
solutions in GR, even in the limit of zero graviton mass. This is
known as the van Dam, Veltman, and Zakharov (vDVZ) discontinuity
\cite{vanDam:1970vg, Zakharov:1970cc}.  Observations such as light
bending in the solar system would exclude the FP theory, no matter
how small the graviton mass is. In 1972, Vainshtein
\cite{Vainshtein:1972sx} proposed a mechanism to avoid this
conclusion; in the small mass limit, the scalar degree of freedom
becomes strongly coupled and the linearized FP theory is no longer
reliable. In this regime, higher order interactions, which are
introduced to remove the ghost degree of freedom, should shield the
scalar interaction and recover GR on sufficiently small scales.

Until recently, it was thought to be impossible to construct a
ghost-free theory for massive gravity that is compatible with
current observations \cite{Creminelli:2005qk, Deffayet:2005ys}.
  Using an effective field theory approach,
  one can show that in  order to avoid the presence of a ghost,
 interactions   have to be chosen in such a way that the equations of motion for the scalar  and vector
component of the St\"uckelberg field  contains no more than two time
derivatives. Recently, it was shown that there is a finite number of
derivative interactions for scalar lagrangians that give rise to
second order differential equations. These are dubbed Galileon terms
because of a symmetry under a constant shift of the scalar field
derivative \cite{Nicolis:2008in}. Therefore, one expects that any
consistent nonlinear completion of FP contains these Galileon terms,
at least in an appropriate range of scales in which the scalar
dynamics can be somehow isolated from the remaining degrees of
freedom; this is  the so-called decoupling limit
\cite{ArkaniHamed:2002sp}. This turns out to be a powerful criterium
for building higher order interactions with the desired properties.
Indeed, following this route, de Rham,  Gabadadze and Tolley
constructed a family of ghost-free extensions to the FP theory,
which reduce to the Galileon terms in the decoupling limit. We refer
to the resulting theory as $\Lambda_3$ massive gravity \cite{dRGT}.
It has been also shown that $\Lambda_3$ massive gravity avoiding the
BD ghost even far from decoupling limit \cite{Hassan:2011zd}. In
this theory, several cosmological solutions have been found, with
particular attention to self-accelerating vacuum solutions which
mimic the $\Lambda$CDM background \cite{Koyama:PRL,
Koyama:solutions, damico, shinji, cosmorefs, Gumrukcuoglu:2011zh,
wyman, wyman:pert, Tasinato:2013rza}. The main goal of this paper is
to study in detail the Hamiltonian structure of perturbations around
these self-accelerating backgrounds based on the approach developed
in \cite{wyman, wyman:pert}. We pay attention to the scalar sector,
where the background fiducial metric choice plays an important role
in characterizing the local dynamics.

Our findings suggest that $\Lambda_3$ massive gravity does act in a
fiducial metric dependent way under certain circumstances. For the
self-accelerating vacuum backgrounds we consider here, there are two
possible behaviours depending on the fiducial metric choice: either
the scalar fluctuations propagate, or there is no propagating scalar
degree of freedom at the linear order in perturbations. In the first
category, we find that the Hamiltonian of the propagating scalar is
unbounded from below, signalling instability regardless of the
choice of the parameters. For the second category of solutions, we
identify the symmetry that eliminates the propagating scalar mode
and show that this symmetry exists when the physical metric and the
fiducial metric have the same form in the background. Due to the
strong coupling behaviour, one should analyse higher order
perturbations to determine stability in this case. A particular
solution with this strong coupling was, indeed, found to be unstable
at a non-linear level \cite{DeFelice:2012mx}.

Finally, to make contact with known solutions in the literature, we
classify some of these space-times, written in different
coordinates, according to these two different behaviours of
perturbations. By taking the decoupling limit of these solutions, we
then discuss the difference between the decoupling theory and the
full theory analysis. It was found that there were regions in the parameter
space where the scalar mode was stable in the decoupling theory
\cite{deRham:2010tw,Koyama:vect1}. On the other hand, vector modes have no dynamics at linear
order in perturbations, but instead acquire dynamics at higher order in fluctuations,
which lead to a Hamiltonian that is unbounded from below \cite{Koyama:vect1}.
At first sight, this result seems inconsistent with our
full theory analysis, where we found that the Hamiltonian is
unbounded from below already at quadratic order if there is a
propagating scalar mode. However, one should remember that the decoupling limit is
not an expansion in field perturbations, but instead a suitable
expansion on the graviton mass $m$. Therefore, some of the features,
such as the instability, in the full theory at linear order in
perturbations may not be captured by the decoupling theory at linear order and
they may emerge at higher order in perturbations. Hence, we conclude that, physically,
our results on the behaviour of perturbations in the two regimes, the decoupling limit and
the full theory, do agree with each other.

\section{Exact Solutions in $\Lambda_3$ Massive Gravity}

Our starting point is the Lagrangian for the $\Lambda_3$ massive
gravity, which has the following form \cite{dRGT}
\begin{align}\label{massive-gravity-Lagrangian}
{\cal{L}}_G=\frac{M_P^2}{2}\sqrt{-g}\left[R-\frac{m^2}{4}{\cal{U}}(g_{\mu\nu},{\cal{K}}_{\mu\nu})\right],
\end{align}
where
\begin{align}\label{K-definition}
\emph{}{\cal{K}}^\mu_{\ \nu}=\delta^\mu_{\ \nu}-\sqrt{\Sigma}^\mu_{\
\nu},\qquad \qquad {\Sigma}^\mu_{\ \nu}\equiv
g^{\mu\alpha}\partial_\alpha \phi^a
\partial_\nu \phi^b \eta_{ab},
\end{align}
and $\phi(x^\mu)$ are the St\"uckelberg fields, which are introduced
to restore the diffeomorphism invariance that was broken by the
choice of fiducial metric $\eta_{ab}$. The mass term ${\cal{U}}$ can
be written in terms of $\Sigma$ as \be {\cal U}= -m^2\left[{\cal
U}_2+\alpha_3\, {\cal U}_3+\alpha_4\, {\cal U}_4\right],
\label{potentialU} \ee with \bea
{\cal U}_2&=&(\tr\mK)^2-\tr (\mK^2),\nonumber \\
{\cal U}_3&=&(\tr \mK)^3 - 3 (\tr \mK)(\tr \mK^2) + 2 \tr \mK^3,\nonumber \\
{\cal U}_4 &=& (\tr \mK)^4 - 6 (\tr \mK)^2 (\tr \mK^2) + 8 (\tr
\mK)(\tr \mK^3) + 3 (\tr \mK^2)^2 - 6 \tr \mK^4 \,,\nonumber \eea
where $m$ has dimension of a mass, while $\alpha_3$ and $\alpha_4$
are dimensionless parameters.

For our purposes it is enough to consider vacuum solutions which
mimic GR backgrounds with a positive cosmological constant. In other
words, we search for vacuum solutions to the Lagrangian
(\ref{massive-gravity-Lagrangian}) which result in a de Sitter space
for the physical metric $g_{\mu\nu}$, supported by non-trivial
configurations of the St\"uckelberg fields $\phi^\mu$. At the
background level, these solutions are indistinguishable from the de
Sitter solution in GR; however, the dynamics of perturbations may
differ. Actually, we find that the latter are
  affected by the choice of fiducial metric in the background level.

To capture this phenomenon we  take the following spherically
symmetric Ansatz for the physical metric, $g_{\mu\nu}$,
\begin{align}\label{metric-background}
ds^2\equiv g_{\mu\nu}dx^\mu
dx^\nu=-b^2(t,r)dt^2+a^2(t,r)(dr^2+r^2d\Omega^2),
\end{align}
with the spherically symmetric St\"uckelberg fields defined as
\be\label{Stuckelberg-fields-background} \phi^0=f(t,r), \qquad
\qquad \qquad \phi^i=g(t,r)\frac{x^i}{r}. \ee A change of frame in
the background metric is accompanied by  a change of the
St\"uckelberg functions $f$, $g$ (see for example the discussion in
\cite{Koyama:solutions}). Due to the above assumptions,  the matrix
$\Sigma^\mu_{\ \nu}$, defined in (\ref{K-definition}), takes the
form
\begin{eqnarray}\label{Sigma}
\Sigma= \left( \begin{array}{cccc}
\frac{\dot f^2-\dot g^2}{b^2} & \frac{\dot ff'-\dot g g'}{b^2} & 0 & 0 \\
\frac{\dot g g'-\dot f f'}{a^2}  & \frac{g'^2-f'^2}{a^2}  & 0 & 0\\
0 & 0 & \frac{g^2}{r^2 a^2} & 0\\
0 & 0 & 0 & \frac{g^2}{r^2 a^2}
\end{array} \right)
\end{eqnarray}
where prime and dot are derivatives with respect to $r$ and $t$,
respectively. This metric choice is particularly helpful to
calculate the square root needed in the Lagrangian definition
(\ref{massive-gravity-Lagrangian})-(\ref{K-definition}). The
equations of motion for $f(t,r)$ and $g(t,r)$ take the following
form \cite{wyman}
\begin{eqnarray}\label{eq.motion-background1}
&&\left(\frac{r^2a^3 P_1}{b \sqrt X}\dot f\right)^.-\left(\frac{r^2a
b P_1}{ \sqrt X}f'\right)'+\mu\left[\left(\frac{r^2 a^2 P_1}{\sqrt
X}+r^2 a^2 P_2\right)^. g'-\left(\frac{r^2 a^2 P_1}{\sqrt X}+r^2 a^2
P_2\right)' \dot g\right]=0,\\\nonumber &&\left(\frac{r^2a^3 P_1}{b
\sqrt X}\dot g\right)^.-\left(\frac{r^2a b P_1}{ \sqrt
X}g'\right)'+\mu\left[\left(\frac{r^2 a^2 P_1}{\sqrt X}+r^2 a^2
P_2\right)^. f'-\left(\frac{r^2 a^2 P_1}{\sqrt X}+r^2 a^2
P_2\right)' \dot f\right]=
  r a^2 b
\left[P'_0+P'_1\sqrt{X}+P'_2W\right]
\end{eqnarray}
where
\begin{eqnarray}\label{XandW}
X=\left(\frac{\dot f}{b}+\mu \frac{g'}{a}\right)^2-\left(\frac{\dot
g}{b}+\mu \frac{f'}{a}\right)^2,\qquad \qquad
W=\frac{\mu}{ab}\left(\dot f g'-\dot g f'\right),
\end{eqnarray}
and $\mu=$sign$\left(\dot f g'-\dot g f'\right)$. The functions
$P_i$ are defined as
\begin{eqnarray}\nonumber
P_0(x)&=&-12-2x(x-6)-12(x-1)(x-2)\alpha_3-24(x-1)^2\alpha_4,
\\\nonumber P_1(x)&=&2(3-2x)+6(x-1)(x-3)\alpha_3+24(x-1)^2\alpha_4,
\\\nonumber P_2(x)&=&-2+12(x-1)\alpha_3-24(x-1)^2\alpha_4,
\end{eqnarray}
and the primes in those functions $P_i$ represent a derivative with
respect to their argument $x=g/(r a)$. The remaining two equations
of motion (with respect to $a$ and $b$) are lengthy and will not be
needed for the arguments below, hence we will not show them.

The equation of motion due to $f$ has a simple solution given by
$g(t,r)=x_0\, r\, a(t,r)$, where $x_0$ is a constant that satisfies
$P_1(x_0)=0$ \cite{wyman}. The last equation for $x_0$ can be
solved, resulting in
\begin{eqnarray}\label{x0}
x_0=\frac{\alpha+3\beta\pm\sqrt{\alpha^2-3\beta}}{3\beta},
\end{eqnarray}
where $\alpha=1+3\alpha_3$ and $\beta=\alpha_3+4\alpha_4$. Notice
that the special case of $\alpha_3=\alpha_4=0$ gives $x_0=3/2$.
Using this solution for $g(t,r)$ we can show that the Einstein
equation is given by \cite{wyman}
\begin{equation}
G^{\mu}_{\;\;\; \nu} = - \frac{1}{2}\;m^2\; P_0(x_0)\;
\delta^{\mu}_{\;\;\; \nu}.
\end{equation}
Thus for self-accelerating solutions that satisfy the condition
$g=x_0\, r \, a$, the functions $a(t,r)$ and $b(t,r)$ are exactly
the same as the scale factor and lapse function in pure GR in
presence of a bare cosmological constant. The remaining function,
$f$, can be obtained from the equation
(\ref{eq.motion-background1}). The non-linearity of the equation
explains why there could be more than one self-accelerating solution
in a given coordinate system.

In the following section we consider perturbations around these
self-accelerating solutions in a general framework, without assuming
any particular choice of coordinates, or any particular profile  for
$f$. In Section \ref{exactsolns}, we present some particular
solutions.

\section{Hamiltonian analysis of perturbations}
In this section, we explore the Hamiltonian structure of scalar
linear perturbations, which only depend on time and radius. In the
notation of the previous Section, we only consider the following
perturbations \bea a(t,r)=a_0(t,r)+\delta a(t,r),\qquad && \qquad
b(t,r)=b_0(t,r)+\delta b(t,r),\\ \nonumber f(t,r)=f_0(t,r)+\delta
f(t,r),\qquad && \qquad g(t,r)=g_0(t,r)+\delta g(t,r), \eea where
the fields with sub-index 0 refer to the background solution.
Actually, the expressions are simplified if one uses the
self-accelerating direction coordinate $\delta\Gamma$, which is
defined as \cite{wyman:pert} \be \delta\Gamma=\delta g-x_0 r \delta
a. \ee The Lagrangian (\ref{massive-gravity-Lagrangian}), to second
order in perturbations, reduces to
\begin{eqnarray}\label{general-2nd-Lag}
{\cal{L}}&=&\delta f\left( A_1  \delta \Gamma + {A_2} \dot {\delta
\Gamma} +{A_3}
 \delta \Gamma' \right)+\delta \Gamma\left({B_1} \delta \Gamma + {B_2} \delta a +{B_3} \dot{\delta a} +{B_4} \delta a'
 +{B_5} \delta b \right)
\\\nonumber&& +\delta a'\left({D_1} \delta b
 +{D_2} \delta a'\right)+\delta b \left({D_3} \delta a''+D_4\delta b+{D_5} \dot {\delta a}\right)+{D_6} \dot{\delta a}^2
 +\delta a\left(E_1  \delta b + E_2\delta a\right),
\end{eqnarray}
where all the capital letters represent functions of $(t,r)$, fixed
by the background solution. We used the background solution for
$g=x_0 r a$, which defines the self-accelerating solutions. The
functions $A_i$, $B_i$ and $E_i$ are associated with  the mass term,
thus have an overall factor of $M^2_{Pl}m^2$, while the $D_i$ arise
from the Hilbert-Einstein piece, hence containing a factor of
$M_{Pl}^2$ only. In what follows we do not need the explicit form of
these functions~\footnote{For their explicit form one can see the
Appendix in \cite{wyman:pert}. Note that we used some integration by
parts.}, except for the relation \be \label{noghost} D_5^2=4D_4D_6,
\ee which ensures the lapse function is a Lagrange multiplier.
Note that there is a special choice of parameters characterised by $\alpha^2 - 3 \beta=0$.
In this case, $A_i=B_i=0$ and there is no propagating scalar mode. In the
rest of this paper, we will not consider this special case.

In order to construct the Hamiltonian, we need the momentum
conjugates of $\delta a$, $\delta b$, $\delta f$ and $\delta
\Gamma$, which read
\begin{eqnarray}
&P_a=B_3\delta \Gamma+D_5\delta b+2 D_6\dot{\delta a},&\qquad \qquad
P_b=0,\\\nonumber &P_f=0,& \qquad  \qquad P_\Gamma=A_2\delta f.
\end{eqnarray}
Before constructing the Hamiltonian in detail, let us explain which
term is the crucial one for the following analysis. It turns out
that $A_2$ is the term that sets the two different behaviours that
we mentioned earlier, and it is related to the fact that the
fiducial metric $\Sigma_{\mu\nu}$ has the same form as the physical
metric $g_{\mu \nu}$: this condition is essentially a choice of
frame. We will come back to this choice of $\Sigma_{\mu\nu}$ later
on, but for now and to explain the different behaviours of the
scalar perturbations, let us consider the Hamiltonian for each case
separately, first for $A_2=0$ and then for $A_2\neq 0$.

\section{Case $A_2=0$: No scalar degrees of freedom}
In this case $P_\Gamma=0$, which results in a constraint, and the
Hamiltonian reads
\begin{eqnarray}
{\cal{H}}&=&\frac{1}{4 {D_6}}(P_a-B_3\delta\Gamma)^2- \delta \Gamma
\left( {B_1} \delta \Gamma  + {B_2}\delta a -{B_4} \delta a'\right)
- {D_2} \delta a'^2-{E_2}  \delta a^2
\\\nonumber&-&  \delta f (A_1\delta \Gamma+ {A_3} \delta \Gamma')- \delta b \bigg(\frac{{D_5}}{2
{D_6}} (P_a-{B_3} \delta \Gamma )+ \left(E_1\delta a +{B_5} \delta
\Gamma +{D_1} \delta a'+{D_3} \delta a''\right)\bigg),
\end{eqnarray}
where we have used (\ref{noghost}) to simplify the expression. By
looking at the above Hamiltonian, it is obvious that $\delta b$ and
$\delta f$ appear linearly, hence their equations of motion are
constraints. Therefore, we end up with the following five primary
constraints
\begin{eqnarray}\label{constraints-A=0}
C_1&=&P_b\, ,\\\nonumber C_2&=&P_f\, ,\\\nonumber
C_3&=&\frac{\partial {\cal{H}}}{\partial \delta f}={ {A_1} \delta
\Gamma+ {A_3} \delta \Gamma'}\, ,\\\nonumber C_4&=&\frac{\partial
{\cal{H}}}{\partial \delta b}=-\frac{{D_5}}{2 {D_6}} (P_a-{B_3}
\delta \Gamma )- \left(E_1\delta a +{B_5} \delta \Gamma +{D_1}
\delta a'+{D_3} \delta a''\right)\, ,\\\nonumber C_5&=&P_\Gamma.
\end{eqnarray}
In addition,  consistency conditions on these primary constraints
lead to an additional secondary constraint, $C_6$, corresponding to
the time
evolution of $C_4$. 
The Poisson algebra of all six constraints results in
\begin{eqnarray}
\nonumber && \{C_j,C_i\}=0\hspace{1cm} j=1,2\hspace{.3cm}
\mathrm{and\ }i\mathrm{ \ arbitrary}\\\nonumber&&
\{C_j,C_i\}\neq0\hspace{1cm} i,j\neq 1,2.
\end{eqnarray}
Therefore, there are two first class constraints, $C_1$ and $C_2$,
and four second class constraints $C_3,C_4,C_5$ and $C_6$, which in
total remove 8 coordinates of the phase space \footnote{Each first
class constraint removes two coordinates of the phase space, while
each second class constraint removes a single coordinate.}.
Therefore, in the case of $A_2=0$, the algebra of constraints
removes all dynamical variables, leaving no propagating scalar
degrees of freedom in the Hamiltonian expanded at quadratic order in
perturbations. Scalar degrees of freedom may acquire non-trivial
dynamics at higher order in perturbations. Indeed it was found that
non-linear perturbations lead to instability \cite{DeFelice:2012mx}.

The absence of a propagating degree of freedom for $A_2=0$ can also
be understood in terms of a new gauge symmetry due to the first
class constraint $C_2$,{ \it i.e.} $P_f=0$. To see this explicitly,
consider the transformation $\delta f \rightarrow \delta f
+\lambda(t,r)$, which induces a change in the Lagrangian
(\ref{general-2nd-Lag}) given by
\begin{eqnarray}
\Delta{\cal{L}}=A_1 \lambda(t,r)\delta\Gamma+A_2
\lambda(t,r)\dot{\delta\Gamma}+A_3\lambda(t,r)\delta\Gamma'=A_2
\lambda(t,r)\dot{\delta\Gamma},
\end{eqnarray}
where we have used the constraint $C_3$ in the last equality. So for
vanishing $A_2$ we obtain $\Delta{\cal{L}}=0$.

Furthermore, a vanishing $A_2$ implies another interesting symmetry
for the fiducial metric $\Sigma_{\mu\nu}$; it presents a same
structure as the physical metric $g_{\mu\nu}$. In order to probe
this statement, let us begin by using equation of motion for $g_0$,
given in (\ref{eq.motion-background1}), which explicitly reads
\begin{eqnarray}\label{eom-g}
\left[\left(r^2 a_0^2 \right)^. f_0'-\left(r^2 a_0^2 \right)' \dot
f_0\right]-2 \mu  r a_0^2 b_0 \left[x_0-\sqrt X_0\right]=0,
\end{eqnarray}
where we have used $g_0=x_0\, r\, a_0$ to restrict ourselves to the
self-accelerating backgrounds. Moreover, using again the
self-accelerating condition, $g_0=x_0\, r\, a_0$, one may write
$A_2=0$ as
\begin{eqnarray}\label{a-condition-1}
a_0^2\dot{f_0}=(r a_0)'\left[(r a_0)' \dot f_0-(r a_0)^. f_0'\right].
\end{eqnarray}
Now by plugging (\ref{a-condition-1}) into (\ref{eom-g}), and using
the definition of $X_0$ from equation (\ref{XandW}), we arrive at
the following equation
\begin{eqnarray}\label{eq}
\frac{1}{b_0^2}\left[\dot f_0^2-x_0^2\big((r
a_0)^.\big)^2-\frac{a_0^2 \dot f_0^2}{\big((r
a_0)'\big)^2}\right]+\frac{2\mu x_0}{b_0}\left[\frac{\dot f_0 (r
a_0)'}{a_0}-\frac{f_0' (r a_0)^.}{a_0}-\frac{a_0 \dot f_0}{(r
a_0)'}\right]+ \left[x_0^2\frac{\big((r
a_0)'\big)^2}{a_0^2}-\frac{f_0'^2}{a_0^2}-x_0^2\right]=0.
\end{eqnarray}
Since the lapse function $b_0$ represents the gauge freedom and it
can be arbitrary, all three brackets in the above equation should
vanish simultaneously. From these conditions, one can show that the
fiducial metric  $\Sigma_{\mu\nu}$ takes the following form
\begin{eqnarray}
\Sigma_{\mu\nu} dx^\mu dx^\nu=-\left(\frac{ a_0 \dot f_0}{(r
a_0)'}\right)^2dt^2+x_0^2a_0^2(dr^2+r^2d\Omega^2),
\end{eqnarray}
which has exactly the same form as the physical metric
(\ref{metric-background}). Note that $a_0$ and $f_0$ are functions
of $(t,r)$.

\section{Case $A_2\neq0$: A single scalar degree of freedom}\label{singlscalsec}
The fact that $A_2\neq0$ implies $P_\Gamma\neq0$, and since
$\dot{\delta \Gamma}$ appears linearly in the Lagrangian we need to
define $\delta f = P_\Gamma/A_2$ to have a well-defined Hamiltonian.
By plugging $\delta f$ in terms of $P_{\Gamma}$ into the
Hamiltonian, we obtain
\begin{eqnarray}\label{ham-A2neq0}
{\cal{H}}&=&-\frac{1}{A_2} P_\Gamma (A_1\delta \Gamma+ {A_3} \delta
\Gamma')+\frac{1}{4 {D_6}}(P_a-B_3\delta\Gamma)^2 -\delta
\Gamma\left( {B_1} \delta \Gamma
 + {B_2}\delta a +{B_4} \delta a'\right)\\\nonumber&-& {D_2} \delta a'^2-{E_2}  \delta a^2-
\delta b \bigg(\frac{{D_5}}{2 {D_6}} (P_a-{B_3} \delta \Gamma )+
\left(E_1\delta a +{B_5} \delta \Gamma +{D_1} \delta a'+{D_3} \delta
a''\right)\bigg).
\end{eqnarray}
We get the four following primary constraints
\begin{eqnarray}\label{constraints-Anot=0}
C_1&=&P_b\, ,\\\nonumber C_2&=&P_f\, ,\\\nonumber
C_3&=&\frac{\partial {\cal{H}}}{\partial \delta b}=-\frac{{D_5}}{2
{D_6}} (P_a-{B_3} \delta \Gamma )- \left(E_1\delta a +{B_5} \delta
\Gamma +{D_1} \delta a'+{D_3} \delta a''\right)\, ,\\\nonumber
C_4&=&P_\Gamma-A_2 \delta f.
\end{eqnarray}
Again, consistency conditions on these primary constraints result in
one additional secondary constraint, $C_5$, which corresponds to the
time evolution of $C_3$. The Poisson algebra of the constraints is
then
\begin{eqnarray}
\nonumber && \{C_j,C_i\}=0\hspace{1cm} j=1\hspace{.3cm} \mathrm{and\
}i\mathrm{ \ arbitrary}\\\nonumber&&
\{C_j,C_i\}\neq0\hspace{1cm}i,j\neq1
\end{eqnarray}
In this case, we have one first class constraint only, $C_1$, and
four second class constraints. Hence we have  2 coordinates in phase
space, corresponding to a single propagating degree of freedom in
the system. It is worth mentioning that in this case $C_2=P_f$ is
not a first class constraint, thus we do not expect the associated
gauge symmetry we had in the previous case.

In this case i.e. $A_2\neq0$, it is interesting to analyse the
stability of the remaining scalar degree of freedom. One can remove
the metric perturbations and their canonical momenta (i.e. $\delta a
$, $\delta b $ and $P_a$) using the constraints $C_3$ and $C_5$, and
obtain the following Lagrangian
\begin{equation}\label{Lag-dec}
{\cal{L}}={A_2} \delta f \dot {\delta \Gamma}+A_1 \delta f \delta
\Gamma+{A_3} \delta f \delta \Gamma'+{\mathcal T}(B_i,D_i,E_i)
\delta \Gamma^2.
\end{equation}
The function ${\mathcal T}(B_i,D_i,E_i)$ is a complicated expression
of the coefficients $B_i$, $D_i$ and $E_i$, which appears as a
consequence of integrating out $P_a$, $\delta a$. The Hamiltonian
derived from the Lagrangian (\ref{Lag-dec}) is given by
\begin{eqnarray}\label{Ham-Gamma}
{\cal{H}}_{\Gamma}&=&-\frac{A_1}{A_2} P_\Gamma \delta
\Gamma-\frac{A_3}{A_2} P_\Gamma \delta \Gamma'-{\mathcal
T}(B_i,D_i,E_i) \delta \Gamma ^2.
\end{eqnarray}
Notice that $P_\Gamma$ appears linearly, implying that this
Hamiltonian is unbounded from below for generic values of the $A_i$,
or equivalently, for arbitrary choices of the self-accelerating
backgrounds solutions. This ``linear" instability is similar to the
instability that appears in higher derivative theories known as
Ostrogradski instability \cite{Ostro}. This instability on its own
is not a bad thing at least classically but this can lead to a
catastrophic instability when this mode couples to healthy degrees
of freedom whose Hamiltonian is bounded from below.

At first sight, this result does not seem to  agree with the
decoupling limit analysis which shows that there is a parameter
space where the Hamiltonian is bounded from below for some
self-accelerating solutions. We will discuss in section \ref{decsec}
this issue; but in order to compare with the decoupling limit
result, we need to know the explicit form of the coefficients that
appear in the Hamiltonian. In the next section we will discuss
explicit solutions for the background functions.

\section{Examples of background solutions}\label{exactsolns}
In this section we will consider three kinds of solutions for the
special case of $\alpha_3=\alpha_4=0$ (a generalisation to any
$\alpha_3$ and $\alpha_4$ is straightforward). These solutions
include those that are previously found in
\cite{Koyama:PRL,Koyama:solutions} and \cite{shinji} (see
\cite{Tasinato:2013rza} for a recent review), as well as a new
solution. The solutions are presented in different coordinates and
we show the existence of a scalar degree of freedom in each
particular fiducial metric choice.

As we have seen, the condition for self-acceleration is $g_0=3 a_0\,
r/2$. This form of $g_0$ leaves no unique solution for $f_0$,
implying that there could be several branches of solutions. In the
literature it has been argued that one branch is defined when
$\Sigma_{\mu\nu}$ has the same symmetries as the physical metric.
However, this property does not hold in all the reference systems as
we will see in what follows. In order to keep the discussion closed
and show enough examples of this coordinate dependence of the
background, it is enough to consider the following backgrounds:
\begin{itemize}
 \item An open-FRWL, with a physical metric given by
 \be\label{metric_frwl}
b_0(t,r)=1, \qquad a_0(t,r)=\frac{\sinh(H\,t)}{4-H^2\,r^2},
 \ee
where $H=  m/2$. As mentioned before, the self-accelerating
backgrounds condition is $g_0=3a_0 r/2$. We show three different
solutions for $f_0$. The first solution, found in
\cite{Koyama:PRL,Koyama:solutions} is given by \bea
 f_0^I&=&\frac{3}{2 H} \left[\text{arctanh}\left(\frac{4 H r }{4-H^2 r^2}\,\,\text{sinh}(H
 t)\right)
 +\text{arctanh}\left(\frac{4+H^2 r^2}{4-H^2 r^2}\,\, \text{tanh}(H t)\right)-\frac{4 H r }{4-H^2 r^2}\,\,\sinh(H t)\right].
 \eea
 The second solution, found in \cite{shinji} but now written in the form of (\ref{metric-background}), is given by
 \be
 f_0^{II}=\frac{3}{2 H}\,\,\frac{4+H^2 r^2}{ 4-H^2
r^2}\,\,\text{sinh}(H t).
 \ee
 Finally, the third and new solution is
 \be
 f_0^{III}=-\frac{3}{H}\,\,\frac{1}{4-H^2 r^2}\,\,
\text{cosh}\left(\frac{H t}{2}\right)\,\, \bigg(-16- H^4 r^4+8 H^2
r^2 \text{cosh}(H t)\bigg)^{\frac{1}{2}}. \ee
\item A flat-FRWL, with a physical metric given by
 \be\label{metric_frwl2}
b_0(t,r)=1 \qquad a_0(t,r)=\frac{1}{2}\,e^{H t},
 \ee
where again $H= m/2$. As mentioned before, the self-accelerating
backgrounds have $g_0=3 a_0 r/2$ and the three solutions equivalent
to those shown above are as follows \bea
 f_0^I&=& \frac{3}{2 H} \left[\text{arctanh}
 \left(\frac{1}{2}\,H\,r\, e^{H t}\right)+\text{arctanh}
 \left(\frac{\left(4+H^2 r^2\right)\,e^{2 H t} -4}{\left(4-H^2 r^2\right)\,e^{2 H t} +4}\right)-\frac{1}{2}\,H\,r\, e^{H t}\right],
\\ f_0^{II}&=&\frac{3}{16 H}\, e^{-H t}\, \bigg(\left(4+H^2
r^2\right)\,e^{2 H t} -4\bigg),
\\f_0^{III}&=&\frac{3}{4 H}\, \, \sqrt{ \left[1+e^{-H t}\right]\times
\left[H^2 r^2e^{2 H t} -4(1+ e^{H t})\right]}. \eea
 \item Conformally flat, with a physical metric given by
 \be\label{metric_conf}
b_0(t,r)=a_0(t,r)=\frac{4}{4+H^2(r^2-t^2)},
 \ee
 where again $H =m/2$. Once again, the spatial part of the St\"uckelberg fields is $g_0=3 a_0 r/2$, while the three solutions become
 \bea
 f_0^{I}&=&\frac{3}{2 H}\left[\text{arctanh}
 \left(\frac{4 H r}{4+H^2 \left(r^2-t^2\right)}\right)+\text{arctanh}\left(\frac{4 H t}{4-H^2 \left(r^2-t^2\right)}\right)
 -\frac{4 H r}{4+H^2 \left(r^2-t^2\right)}\right],\\
 f_0^{II}&=&\frac{6 t}{4+H^2(r^2-t^2)},\\
 f_0^{III}&=&\frac{6\sqrt{H^2t^2-4}}{H(4+H^2(r^2-t^2))}.
 \eea
\\From the last expression, we see that solution III is valid for times larger than the Hubble scale, {\it i.e.} $t\geq 1/H$.

 \end{itemize}

In order to exhibit the different behaviours of scalar
perturbations, it is useful to write the explicit form of $A_2$,
which is given by
\begin{eqnarray}
A_2=-4\left[\frac{3\dot{f_0}}{2b_0 W_0}-\mu\frac{(r
a_0)'}{a_0}\right],
\end{eqnarray}
where as mentioned before index $0$ shows the background variables.
 From this coefficient, one can determine if there is a propagating
d.o.f. using the analysis of the previous Sections. Table
\ref{table} summarises the three solutions (I,II and III) in the
three different frames we have written above (open-FRWL, flat-FRWL
and conformally flat). It is interesting to notice that solution II,
found in \cite{shinji}, only has strong coupling in scalar sector in
the open-FRWL frame, in agreement with \cite{Gumrukcuoglu:2011zh}.
Moreover, solution I, found in \cite{Koyama:PRL, Koyama:solutions},
does propagate a scalar d.o.f. in all three frames given here.
Finally, the new solution (III) in the conformal frame does not
propagate a scalar mode at linear order in perturbations.

\begin{table}[ht]
\begin{center}
    \begin{tabular}{ | c | c | c | c  |}
    \hline
    Background solution & I & II & III \\ \hline
    open-FRWL & $A_2 \neq 0$& $A_2 = 0$ & $ A_2 \neq 0$ \\ \hline
    flat-FRWL & $A_2 \neq 0$ & $A_2 \neq 0$ & $A_2 \neq 0$  \\ \hline
    conformally flat  & $A_2 \neq 0$ & $A_2 \neq 0$ & $A_2 = 0$ \\
    \hline
    \end{tabular}
    \caption{Three self-accelerating solutions with the corresponding $A_2=0$ condition in three different background coordinate choices.
     Solutions which satisfy $A_2=0$ have no propagating scalar d.o.f. at linear order in perturbations, whereas
    solutions with $A_2\neq0$ propagate a single scalar mode.}\label{table}
\end{center}
\end{table}

\section{Decoupling limit}\label{decsec}
In this section, we discuss the decoupling limit case and clarify
the difference between the decoupling limit theory and the full
theory analysis. The decoupling limit is defined as $m\rightarrow
0$, $M_{pl}\rightarrow \infty$ with $\Lambda_3 \equiv M_{pl}m^2$
fixed. In order to take this limit we need to normalise the fields
in the following way:
$$\delta a\rightarrow M_P^{-1}\delta a,\qquad \delta b\rightarrow
M_P^{-1}\delta b,\qquad \delta f\rightarrow\Lambda_3^{-1} \delta
f\quad \mathrm{ and }\quad \delta g\rightarrow\Lambda_3^{-1} \delta
g.$$ Under this rescaling, the Lagrangian (\ref{general-2nd-Lag})
reads
\begin{eqnarray}\label{Lag-decoupling}
{\cal{L}}&=&{D_1}\delta b \delta a'+{\tilde{D}_2} \delta
a'^2+{\tilde{D}_3}\delta b \delta a''+\tilde{D}_4\delta
b^2+{\tilde{D}_5} \delta b\dot {\delta a}+{\tilde{D}_6} \dot{\delta
a}^2\\\nonumber&+&m^2\left[\tilde{E}_1\delta b\delta
a+\tilde{E}_2\delta
a^2\right]\\\nonumber&+&\frac{1}{m^2}\left[\tilde{A}_1 \delta f
\delta \Gamma+{\tilde{A}_2} \delta f \dot {\delta
\Gamma}+{\tilde{A}_3} \delta f \delta \Gamma'+{\tilde{B}_1} \delta
\Gamma^2\right]\\\nonumber &+&\left[{\tilde{B}_2}\delta \Gamma
\delta a+{\tilde{B}_3} \delta \Gamma \dot{\delta a}+{\tilde{B}_4}
\delta \Gamma \delta a'+\tilde{B}_5 \delta \Gamma\delta b \right],
\end{eqnarray}
where we have pulled out all the $m$ and $M_{Pl}$ dependence from
the capital functions $A_i$, $B_i$, $D_i$ and $E_i$ (leaving
expressions  with a tilde) and also used $M_{Pl}=\Lambda_3/m^2$ to
write everything in terms of $m$ and $\Lambda_3$. The decoupling
limit is then obtained by the $m\rightarrow 0$, with $\Lambda_3$
fixed. It is worth mentioning that the first line comes from pure
Einstein Hilbert action and the three other lines come from the mass
term.

To go further we need to know the behaviour of coefficients in the
$m\rightarrow 0$ limit. For this purpose we use the decoupling limit
of the background solutions given in the previous section. For the
self-accelerating solutions, the Hubble parameter $H$ is
proportional to $m$. Thus in the decoupling limit we take the limit
$Ht, Hr \ll 1$. In order to have a Minkwoski spacetime in this
limit, we use the conformal metric frame when taking this limit. We
should note that the decoupling limit of the solution III is not
well defined, because $f_0^{III}$ becomes imaginary in this limit.
This is a special solution where there is no propagating degree of
freedom, thus it does not contradict the decoupling limit analysis
of \cite{deRham:2010tw, Koyama:vect1}, which showed that the
self-accelerating solution in the decoupling limit propagates a
single scalar mode unless $\alpha^2 - 3 \beta=0$. On the other hand
solutions I and II have the same decoupling limit solutions
\cite{Koyama:vect1}. Note that solution II has a propagating scalar
mode in the conformally flat frame, in contrast to the same solution
in the open-FRWL frame where the full theory has no propagating
scalar degree of freedom. Again this is not a contradiction, as the
decoupling limit is not well defined in the open-FRW frame. In the
decoupling limit, the background solutions are given by
 \begin{eqnarray}\label{decoupling-background}
a_0=b_0=1-\frac{H^2}{2} \left(r^2-t^2\right),\qquad f_0=x_0^\pm
 t,\qquad g_0=x_0^\pm  r,\qquad
H^2=\frac{m^2}{3}\frac{\alpha\mp
2\sqrt{\alpha^2-3\beta}}{\left(3\alpha\mp
\sqrt{\alpha^2-3\beta}\right)^2},
\end{eqnarray}
with $x_0$, $\alpha$ and $\beta$ defined in and below (\ref{x0}). It
is possible to show that in $m\rightarrow 0$ limit the relevant
terms come from the first and third line in (\ref{Lag-decoupling}).
If one then describes the scalar mode in the usual way in the
decoupling theory ({\it i.e.} $\phi^\mu=x^\mu-\partial^\mu\pi$,
where $\pi$ is the scalar mode, and is equivalent to $\delta f=-\dot
\pi$ and $\delta\Gamma=\pi'$) then the scalar Lagrangian in the
decoupling limit becomes  \cite{deRham:2010tw,Koyama:vect1}
\begin{eqnarray}\label{lag-pi}
{\cal{L}}_{kin.}=\pm 3 \sqrt{\alpha^2-3\beta}\Lambda_3^2 \left(
\frac{H}{m} \right)^2 \pi\Box \pi.
\end{eqnarray}
The associate Hamiltonian is
\begin{eqnarray}\label{Ham-pi}
{\cal{H}}_{\pi}= \pm \Big( \frac{
1}{\sqrt{\alpha^2-3\beta}}\,\frac{12}{\Lambda_3^2} \left(
\frac{m}{H} \right)^2\, P_\pi^2 + 3
\sqrt{\alpha^2-3\beta}\Lambda_3^2 \left( \frac{H}{m} \right)^2
\pi'^2 \Big),
\end{eqnarray}
which implies that the scalar perturbations are stable (unstable)
for the $+$ ($-$) branch \cite{deRham:2010tw,Koyama:vect1}. For the
special case $\beta=0$, which includes $\alpha_3=\alpha_4=0$, the
$+$ branch of solutions disappears and there is always a ghost.

At first sight, this result seems inconsistent with our previous
full theory analysis, where we found that the Hamiltonian is
unbounded from below for all the self-accelerating solutions if $A_2
\neq 0$. However, one should remember that the decoupling limit is
not an expansion in field perturbations, but instead a suitable
expansion on the graviton mass $m$ (keeping only the leading terms
to a finite  scale $\Lambda_3$). Therefore, some of the features,
such as the instability, in the full theory at linear order in
perturbations may not be captured by the decoupling theory at linear
order. However, they may emerge at higher order in perturbations in
the decoupling limit. This interpretation is supported by previous
findings on the dynamics of vector degrees of freedom in the
decoupling limit of massive gravity \cite{Koyama:vect1}. In these
papers, it was shown that vector modes have no dynamics at linear
order in perturbations, but instead acquire dynamics at higher order
in fluctuations, which in turn, lead to a Hamiltonian that is
unbounded from below  -- exactly as we find in the full theory
analysis. Hence, our results on the behaviour of perturbations in
the two regimes, the decoupling limit and the full theory,
physically agree with each other. Finally, we conclude that the
self-accelerating solutions are generically unstable to linear
perturbations, which together with other problems \cite{deser}, put
some pressure on the viability of this model to explain
observations.

\begin{acknowledgments}
NK acknowledges bilateral funding from the Royal Society and the
South African NRF which supported this project. GN is supported by
the grants PROMEP/103.5/12/3680 and CONACYT/179208. NK and GN also
thank the Institute of Cosmology and Gravitation for its hospitality
during their visits. KK is supported by STFC grant ST/H002774/1 and
ST/K0090X/1, the European Research Council and the Leverhulme trust.
GT is supported by an STFC Advanced Fellowship ST/H005498/1.
\end{acknowledgments}

\end{document}